\documentclass[12pt]{article}
\usepackage{amssymb,amsmath} 
\usepackage{graphicx}
\usepackage{epsfig} 
\usepackage{latexsym}
\usepackage{psfrag}   
\usepackage{subfigure}  
\usepackage{booktabs}  
\usepackage{braket}
\usepackage{mathtools}
\usepackage{protosem}
\usepackage{wasysym}
\usepackage[numbers,sort&compress]{natbib}
\usepackage{textcomp}
\usepackage[utf8]{inputenc}

\usepackage{mathrsfs}
\usepackage[inner=2.9cm,outer=2cm]{geometry}
\usepackage[toc]{appendix}
\usepackage{color,soul} 
\usepackage{datetime}
\usepackage[
      colorlinks=true,
      linkcolor=darkblue,  
      urlcolor=black,    
      filecolor=blue,     
      citecolor=darkgreen, 
      linktocpage=true,
      pdfstartview=FitV,
      bookmarksopen=true    
      ]{hyperref}

\definecolor{darkblue}{rgb}{0.2, 0, 0.8}
\definecolor{darkgreen}{rgb}{0.2, 0.71, 0}

\numberwithin{equation}{section}

\newcommand{\bea}{\begin{eqnarray}}
\newcommand{\eea}{\end{eqnarray}}
\newcommand{\ba}{\begin{eqnarray}}
\newcommand{\ea}{\end{eqnarray}}

\newcommand{\beq}{\begin{equation}}
\newcommand{\eeq}{\end{equation} }
\newcommand{\beqa}{\begin{eqnarray}}
\newcommand{\eeqa}{\end{eqnarray}}
\newcommand{\beqar}{\begin{eqnarray*}}
\newcommand{\eeqar}{\end{eqnarray*}}

\renewcommand{\href}[2]{#2}

\newenvironment{changemargin}[2]{%
\begin{list}{}{%
\setlength{\topsep}{0pt}%
\setlength{\leftmargin}{#1}%
\setlength{\rightmargin}{#2}%
\setlength{\listparindent}{\parindent}%
\setlength{\itemindent}{\parindent}%
\setlength{\parsep}{\parskip}%
}%
\item[]}{\end{list}}

\begin{document}  


\begin{titlepage}


\vspace*{1.2cm}

\begin{center}
{\Large {\bf Non-Abelian bubbles in microstate geometries}} \\

\vspace*{1.2cm}
\renewcommand{\thefootnote}{\alph{footnote}}
{\sl\large Pedro F.~Ram\'{\i}rez$^{\text{\eighthnote}}$ }\footnotetext{$^{\text{\eighthnote}}$ {\tt p.f.ramirez [at] csic.es}}

\setcounter{footnote}{0}
\renewcommand{\thefootnote}{\arabic{footnote}}

\vspace{1.5cm}

{\it Instituto de F\'{\i}sica Te\'orica UAM/CSIC\\
C/ Nicol\'as Cabrera, 13--15,  C.U.~Cantoblanco, E-28049 Madrid, Spain}\\ 

\vspace{0.3cm}

{\it Institut de Physique Th\'eorique, Universit\'e Paris Saclay, CEA\\ CNRS, F-91191 Gif-sur-Yvette, France} \\

\bigskip

\end{center}

\vspace*{0.1cm}

\begin{abstract}  
\begin{changemargin}{-0.95cm}{-0.95cm}
We find the first smooth microstate geometries with non-Abelian fields. The solutions constitute an extension of the BPS three-charge smooth microstates. These consist in general families of regular supersymmetric solutions with non-trivial topology, i.e. \emph{bubbles}, of $\mathcal{N}=1$, $d=5$ Super-Einstein-Yang-Mills theory, having the asymptotic charges of a black hole or black ring but with no horizon. The non-Abelian fields make their presence at the very heart of the microstate structure: the physical size of the bubbles is affected by the non-Abelian topological charge they carry, which combines with the Abelian flux threading the bubbles to hold them up. Interestingly the non-Abelian fields carry a set of adjustable continuous parameters that do not alter the asymptotics of the solutions but modify the local geometry. This feature can be used to obtain a \emph{classically infinite} number of microstate solutions with the asymptotics of a single black hole or black ring.
\end{changemargin}
\end{abstract} 

\end{titlepage}

\setcounter{tocdepth}{2}
{\small
\setlength\parskip{-0.5mm} 
\noindent\rule{15.7cm}{0.4pt}
\tableofcontents
\vspace{0.6cm}
\noindent\rule{15.7cm}{0.4pt}
}

\section{Introduction}


The construction and study of smooth microstate geometries in supergravity theories has become a fruitful area of research since the pioneering work, more than a decade ago, of Bena and Warner \cite{Bena:2005va} and independently of Berglund, Gimon and Levi \cite{Berglund:2005vb}, where a strategy to obtain ample families of microstate geometries was given, generalizing earlier results \cite{Lunin:2001jy, Lunin:2002iz, Mathur:2003hj, Lunin:2004uu, Giusto:2004id, Giusto:2004ip, Giusto:2004kj}. This kind of solutions can be roughly described as a black hole configuration in which the horizon and its interior have been replaced by some complicated, although smooth horizonless geometry while keeping the rest of the field configuration looking like the unmodified solution. Any solution with such remarkable properties is interesting \emph{per se}, although it is in the context of the \emph{fuzzball proposal} \cite{Mathur:2005zp} in which these configurations acquire their greatest significance.

The proposal originated as a possible solution to the information paradox and conjectures that the entropy of a black hole has its microscopic origin in the degeneracy of a quantum bound state, the fuzzball. In this picture, the classical black hole would provide an effective description of the system, that would consist in a quantum \emph{ensamble of geometries}. These microstate geometries, when considered individually, would correspond to string theory configurations with unitary scattering and hopefully a subset of these states might be captured as smooth horizonless supergravity solutions. Since the proposal suggests a modification at the horizon scale, such geometries should have the same asymptotics as the black hole.

This conjecture opened a whole program in the quest to construct smooth microstate geometries in theories of supergravity. Much progress has been made in this direction and vast classes of such solutions have already been described in the literature, see \cite{Bena:2007kg, Skenderis:2008qn, Balasubramanian:2008da, Chowdhury:2010ct, Bena:2013dka} and references therein. The direct identification of these configurations as representing typical microstates of a particular black hole is generally unclear due to the absence of a description in terms of a dual CFT. However very recently this identification has been performed for a particular type of configurations known as \emph{superstrata}, constituting a major achievement of the fuzzball program \cite{Bena:2016ypk}. Nevertheless, even though general microstate geometries lack of this identification, they are still very useful in providing valuable information about the physics of black holes in string theory, see for instance \cite{Mathur:2012jk, Bena:2012zi, Bena:2015lkx, Bena:2015dpt, Roy:2016zzv}.

Typically these are described in five and six dimensions, in the context of supergravity coupled to Abelian matter multiplets or pure supergravity. In the present work we perform the inclusion of non-Abelian degrees of freedom for the first time. The reason why this class of microstate geometries has remained unexplored so far seems to be clear: the construction of explicit analytic non-Abelian solutions in five- and six-dimensional supergravity theories has become accessible only in the last few months \cite{Meessen:2015enl, Ortin:2016bnl, Cano:2016rls}. The solutions that we present here constitute a non-Abelian extension of the BPS three-charge smooth geometries described in \cite{Bena:2007kg}. We work in $\mathcal{N}=1$, $d=5$  Super-Einstein-Yang-Mills (SEYM) theories. One can think of these theories as an extension of the five-dimensional STU model of supergravity, that describes a supergravity multiplet coupled to two Abelian vector multiplets. SEYM theories are then obtained by consistently coupling the STU model to a set of additional vector multiplets that transform under the local action of a non-Abelian group\footnote{One can consider as well the introduction of additional Abelian vector multiplets.}. Although this nomenclature might seem unfamiliar in the literature of microstate geometries, in fact the underlying theory where this solutions are constructed is quite frequently the STU model: five-dimensional three-charge configurations are naturally described in this framework.

The procedure by which non-Abelian microstate geometries are found has a similar structure than that of the Abelian case, but requires the introduction of some modifications. Just like in the case of supersymmetric solutions of STU supergravity, the construction of BPS configurations satisfying the equations of motion of SEYM theory relies on the specification of a reduced set of \emph{seed functions} defined in $\mathbb{R}^3$. In the case of the familiar STU model, these are simply harmonic functions that satisfy certain differential equations whose integrability condition is the Laplace equation. The SEYM procedure conserves these harmonic functions and introduces a new set of seed functions satisfying the \emph{covariant} version of these differential equations.

We find that the bubbling equations, which determine the size of the bubbles leading to physically sensible geometries, contain a new contribution that appears standing next to the magnetic fluxes threading the bubbles, see (\ref{eq:bubbling}). This new term can be given a physical interpretation in terms of the topological charge, or instanton number, associated to the endpoints of the bubble of a non-Abelian instanton that builds up the vector fields. As a consequence it should be possible to have stable bubbles without some magnetic fluxes placed on them or, inversely, a bubble can collapse even though the fluxes are non-zero.

Another interesting peculiarity introduced by the non-Abelian fields is that the solution depends on a set of continuous parameters that can be modified with no apparent restriction whose influence is only local, i.e. their modification does not change any of the asymptotic charges. This is a shocking feature that allows the construction of huge amounts of microstate geometries with the same topology for a unique black hole, and its proper interpretation requires further study.

Having said that, let us start talking about the details of non-Abelian microstate geometries. We give a general description of the solutions that can be found using our generating technique in Section \ref{sec:2}. In Section \ref{sec:bubbling} we describe how this method can be utilized for the construction of smooth horizonless solutions. We conclude in Section \ref{sec:final} with some comments about the results and discuss future directions. In Appendix \ref{sec:theory} we give a brief summary of $\mathcal{N}=1$, $d=5$ SEYM theories, describing its matter content and its action. Appendix \ref{sec:recipe} contains the solution generating technique written in a \emph{step-by-step} language.




\section{Supersymmetric solutions of $\mathcal{N}=1$, $d=5$  Super-Einstein-Yang-Mills}
\label{sec:2}

A technique to construct supersymmetric timelike solutions with a spacelike isometry in these theories was recently developed in \cite{Meessen:2015enl}, where it was used to describe the first non-Abelian analytic black holes in five dimensions\footnote{A method for the systematic construction of null solutions and some explicit examples describing black strings and regular string-monopoles are also given in that reference.}. This method has also been used in \cite{Ortin:2016bnl} to find non-Abelian generalizations of the Emparan-Reall black ring solution, \cite{Emparan:2001wn}, and the BMPV rotating black hole, \cite{Breckenridge:1996is}. In the simplest settings, the configurations can be roughly interpreted as three-charge Abelian solutions on top of which we place a non-Abelian instanton that, interestingly, does not produce any change on the mass of the solution while it reduces its entropy.

The solutions of $\mathcal{N}=1$, $d=5$ SEYM\footnote{See Apendix \ref{sec:theory} for a brief description of the theory.} are specified by the form of the metric $ds^2$, the vector fields $A^I$ and the scalars $\phi^x$. The indices labeling the vectors take values in $\{ I,J,\ldots=0,\ldots,5 \}$, with the Abelian sector contained in the first values $\{i,j,\ldots=0,1,2 \}$ and the non-Abelian sector in the last three $\{ \alpha,\beta,\ldots=3,4,5 \}$. We make a continuous use of this division in two sectors through the text. The scalars are conveniently codified in terms of a set of functions $h_I$ labeled with the same indices than the vectors, such that $\phi^x \equiv h_x/h_0$. We also define the functions of the scalars with upper indices as

\begin{equation}
h^I \equiv 27 C^{IJK} h_I h_J \, , \hspace{1.5cm} h^I h_I = 1 \, ,
\end{equation}

\noindent
where $C^{IJK} = C_{IJK}$ is a constant symmetric tensor that characterizes the supergravity theory. We work on the SU$(2)$-gauged ST$[2,6]$ model, that contains $n_v=5$ vector multiplets and, as we mentioned in the introduction, can be understood as a non-Abelian extension of the STU model. This model is characterized by a constant symmetric tensor with the following non-vanishing components
\begin{equation}
C_{0xy}= \tfrac{1}{6}\eta_{xy}\, ,
\mbox{where}
\,\,\,\,\,
(\eta_{xy}) = \mathrm{diag}(+-\dotsm -)\, ,
\,\,\,\,\,
\mbox{and}
\,\,\,\,\,
x,y=1,\cdots,5\, .
\end{equation}

In \cite{Bellorin:2007yp} it was shown that timelike supersymmetric solutions of this theory are of the form
\begin{eqnarray}
\label{eq:themetric}
ds^{2} 
&=& 
f^{\, 2}(dt+\omega)^{2}
-f^{\, -1}d\hat{s}^{2}\, , \\
& & \nonumber \\
\label{eq:vectorfields}
A^{I} 
& = &
-\sqrt{3}h^{I} f (dt +\omega) +\hat{A}^{I}\, ,  
\end{eqnarray}

\noindent 
where $d\hat{s}^2$ is a four-dimensional hyperK\"ahler metric and the rest of elements that appear in this decomposition are defined on this four-dimensional space. These elements satisfy the system of \emph{BPS equations}:
\begin{eqnarray}
\hat{F}^I &=& \star_4 \hat{F}^I \, , \\
& & \nonumber \\
\label{eq:BPS}
\hat{\mathfrak{D}}^2 \left( h_I / f \right) &=& \tfrac{1}{6} C_{IJK} \hat{F}^J \cdot \hat{F}^K \, , \\
& & \nonumber \\
d\omega + \star_4 d\omega &=& \tfrac{\sqrt{3}}{2} (h_I/f) \hat{F}^I \, .
\end{eqnarray}

\noindent
Here $\star_4$ is the Hodge dual in the four-dimensional metric $d\hat{s}^2$ and $\hat{F}^I$ is the field strength of the vector $\hat{A}^I$
\begin{equation}
\hat{F}^{I}{}_{\mu\nu}=2\partial_{[\mu}\hat{A}^{I}{}_{\nu]}+\hat{g}f_{JK}{}^{I}\hat{A}^{J}{}_{\mu}\hat{A}^{K}{}_{\nu} \, ,
\end{equation}

\noindent
where $f_{IJ}\,^{K}$ are only non-vanishing when the indices take values in the non-Abelian sector, in which case they are the structure constants of $SU(2)$, $f_{\alpha\beta}\,^{\gamma}=\varepsilon_{\alpha\beta\gamma}$.

Some words about notation are necessary. Notice that we use \emph{hats} to distinguish objects that are defined in four spatial dimensions. For example, $A^I$ is used to represent the five-dimensional physical vectors and $\hat{A}^I$ is a vector in the four-dimensional hyperK\"ahler space. In a few lines we will introduce another collection of objects that are labeled with \emph{inverse hats} and that are defined in three-dimensional Euclidean space. In particular we define the vectors $\breve{A}^I$. We use all these vectors to define covariant derivatives in five, four and three dimensions for objects with upper and lower vector indices. For example the four-dimensional covariant derivatives are defined by
\begin{equation}
\hat{\mathfrak{D}} h^I = d h^I + \hat{g} f_{JK}\,^{I} \hat{A}^J h^K \, , \hspace{2cm} \hat{\mathfrak{D}} h_I = d h_I + \hat{g} f_{IJ}\,^{K} \hat{A}^J h_K \, .
\end{equation}

The system of BPS equations can be drastically simplified under the assumption that the solutions admit a global spacelike isometry along a compact direction \cite{Meessen:2015enl}. Then the mathematical objects that build up the physical fields can be further decomposed in terms of elements defined in three dimensional flat space in the following manner
\begin{eqnarray}
\label{eq:GHmetric}
d\hat{s}^{2}
&=&
H^{-1} (d\varphi +\chi)^{2}
+H dx^{r}dx^{r}\, ,\,\,\,\, r=1,2,3\, , \\
& & \nonumber \\
\label{eq:instantondec}
\hat{A}^{I}
& = &
-2\sqrt{6} \left[-H^{-1}\Phi^{I} (d\varphi+\chi)+\breve{A}^{I} \right]\, , \\
& & \nonumber \\
\label{eq:hIf}
h_{I}/f 
&=& 
L_{I}+8C_{IJK}\Phi^{J}\Phi^{K}H^{-1}\, , \\
& & \nonumber \\
\label{eq:omegahat}
\omega 
& = & 
\omega_{5}(d\varphi+\chi) +\breve{\omega}\, ,
\end{eqnarray}

\noindent
where $\varphi$ is a coordinate adapted to the direction of the isometry. Substituting back these expressions in the BPS system of equations, we obtain the conditions that $H, \chi, \Phi^I, \breve{A}^I, L_I, \omega_5$ and $\breve{\omega}$ need to satisfy
\begin{eqnarray}
\label{eq:Hequation}
\star_{3}dH 
& = &
d\chi  \, ,
\\
& & \nonumber \\
\label{eq:PhiIequation} 
\star_{3}\breve{\mathfrak{D}} \Phi^{I}
& = &
\breve{F}^{I}\, ,
\\
& & \nonumber \\
\label{eq:LIequation}
\breve{\mathfrak{D}}^{2}L_{I} 
& = &
\breve{g}^{2} f_{IJ}{}^{L}f_{KL}{}^{M}\Phi^{J}\Phi^{K}L_{M}\, ,  
\\
& & \nonumber \\
\label{eq:omegaequation2}
\star_{3} d \breve{\omega}
&=&
H dM-MdH
+3\sqrt{2} ( \Phi^{I} \breve{\mathfrak{D}}L_{I}
-L_{I}\breve{\mathfrak{D}} \Phi^{I} )
\, , \\
& & \nonumber \\
\label{eq:omega5}
\omega_5
&=&
M+16\sqrt{2} H^{-2} C_{IJK} \Phi^{I} \Phi^{J} \Phi^{K}
+3\sqrt{2} H^{-1} L_I \Phi^{I} \, ,
\end{eqnarray}

\noindent
where $M$ is just a harmonic function in $\mathbb{E}^3$, i.e. $\nabla^2 M =0$.

Equations (\ref{eq:Hequation}), (\ref{eq:PhiIequation}) and (\ref{eq:LIequation}) in the Abelian sector imply that $H$, $\Phi^i$ and $L_i$ are just harmonic functions on $\mathbb{E}^3$. Once these are specified it is straightforward to find the 1-forms $\chi$ and $\breve{A}^i$.

In the non-Abelian sector (\ref{eq:PhiIequation}) is the Bogomol'nyi equation \cite{Bogomolny:1975de}, which is non-linear and hard to solve in general. Fortunately this system, that describes a non-Abelian monopole in Yang-Mills-Higgs theory, has been studied by many authors and the space of solutions available in the bibliography is rich enough for the purposes of our work. 

Equation (\ref{eq:LIequation}) in the non-Abelian sector is easily solved if we choose $L_{\alpha}\propto \Phi^{\alpha}$ or just $L_\alpha=0$. However none of these choices is completely satisfying if one pursues the construction of general smooth horizonless geometries. If one takes $L_{\alpha}\propto \Phi^{\alpha}$ then there are some potential restrictions on the space of possible $\Phi^i$ that can result in smooth geometries. We will need to find a more general solution.

Finally, (\ref{eq:omegaequation2}) can always be solved if its integrability condition is satisfied. This condition gives a set of algebraic equations, which in this context are known as \emph{bubbling equations}, that impose restrictions on the distance between the different centers of the solution (the points were the seed functions are singular). Then, of course, one has to integrate explicitly equation (\ref{eq:omegaequation2}) to obtain $\breve{\omega}$.

In summary, we have described a procedure to construct supersymmetric timelike solutions in terms of a set of \emph{seed functions} defined on three-dimensional flat space: $H, \Phi^I, L_I$ and $M$.

\section{Smooth bubbling geometries in SEYM supergravity} 
\label{sec:bubbling}

Smooth microstate geometries are defined as horizonless, regular field configurations without any brane sources but with the asymptotic charges of a black hole. At a technical level this statement implies several conditions that we shall address in the following subsections, being perhaps the most important of those the requirement of working with manifolds with non-trivial topology\footnote{By this we mean that they describe non-contractible spaces.}. This fact can be roughly understood from the fact that the existence of non-trivial cycles allows for the presence of measurable asymptotic charges without the introduction of localized brane sources. See for instance \cite{Bena:2007kg} for a detailed discussion about this topic.

%
%

The systematic procedure for finding solutions described in the previous section can naturally accommodate ambipolar Gibbons-Hawking spaces, which have just the right properties for these purposes. Let us start with a brief description of these manifolds.


\subsection{Ambipolar Gibbons-Hawking spaces}
\label{sec:ambipolar}

Much of the very interesting physics exhibited by these solutions is related to the use of ambipolar Gibbons-Hawking spaces, which are a particular example of ambipolar hyperK\"ahler manifolds \cite{Niehoff:2016gbi}. These have the form of a $U(1)$ fibration over a $\mathbb{R}^3$ base, with the fiber collapsing to a point at a finite collection $X =\{ \vec{x}_a \vert a=1,\ldots,n \}$ of points in $\mathbb{R}^3$ which we will call \emph{centers}. Any path in the base manifold connecting two centers, $\gamma_{ab}$, defines a non-contractible 2-cycle through the inclusion of the $U(1)$ fiber, $\Delta_{\gamma_{ab}}$. A different path $\gamma'_{ab}$ between the same centers describes an homologically equivalent 2-cycle $\Delta_{\gamma'_{ab}} \simeq \Delta_{\gamma_{ab}}$. We will denote any of the equivalent 2-cycles simply as $\Delta_{ab}$.

These spaces have the metric
\begin{equation}
d\hat{s}^2 = H^{-1} (d\varphi +\chi)^{2}+H \left[ dr^2 + r^2 \left( d\theta^2 + sin^2 \theta d\psi^2 \right) \right] \, , \hspace{1cm} \star_3 dH=d\chi \, ,
\end{equation}

\noindent
with the angular coordinates taking values in $\theta \in [0, \pi)$, $\psi \in [0, 2\pi)$, $\varphi \in [0, 4\pi)$. $H$ is a harmonic function on $\mathbb{E}^3$ of the form

\begin{equation}
H=\sum_a \frac{q_a}{r_a} \, , \hspace{1.5cm} \text{with} \hspace{0.5cm}  r_a \equiv \vert \vec{x} - \vec{x}_a \vert \, , \hspace{0.5cm} \vec{x}_a \in X \, ,
\end{equation}

\noindent
while the 1-form $\chi$ plays the role of local connection of the fiber bundle and can be written as
\begin{equation}
\chi = \sum_a q_a cos \theta_a d\psi_a \, ,
\end{equation}
where $\theta_a$ and $\psi_a$ are coordinates on a spherical frame centered in $\vec{x}_a$.

Although $H$ is singular when evaluated at the centers it is straightforward to check that if all $q_a$, aka Gibbons-Hawking charges, are integers then the metric remains regular at these points\footnote{When $\vert q_a \vert \neq 1$ there is an orbifold singularity at $\vec{x}=\vec{x}_a$, but we will not worry about it since these singularities are innocuous in the context of string theory.}. Indeed under the redefinition of the radial coordinate $\rho_a=2 \sqrt{r_a}$ we find that locally
\begin{equation}
d\hat{s}^2 \vert_{\rho_a \rightarrow 0} \sim d\rho_a^2 + \rho_a^2 d\Omega^2_{(3)/q_a} \, ,
\end{equation}
being $d\Omega^2_{(3)/q_a}$ the standard metric on $S^3/\mathbb{Z}_{\vert q_a \vert}$. Asymptotically the manifold is also of this form, $d\hat{s}^2 \vert_{\rho \rightarrow \infty} \sim d\rho^2 + \rho^2 d\Omega^2_{(3)/Q}$, with the orbifold given in this case by $S^3/\mathbb{Z}_{\vert Q \vert}$, being $Q \equiv \sum_a q_a$.

Physically, smooth bubbling geometries are claimed to represent microstate configurations of some particular black hole, being both solutions indistinguishable asymptotically. Therefore we are interested in having the ambipolar Gibbons-Hawking space asymptotic to $\mathbb{R}^4$, which we can achieve imposing $Q=1$. This condition requires that some of the Gibbons-Hawking charges be negative, and therefore the function $H$ interpolates between $-\infty$ and $+\infty$. Each negatively charged center is surrounded by a connected open region with $H<0$, whose boundary is a surface where $H$ vanishes.

Then the signature of the metric interpolates between $\left(++++\right)$ and $\left(----\right)$, being clearly ill-defined at the surfaces where $H=0$. It is this characteristic what renders this space be ambipolar. This harmful properties, however, can be made compatible with having a smooth five-dimensional supergravity solution due to the presence of both, the conformal factor $f^{-1}$ multiplying $d\hat{s}^2$ and the additional terms in the full metric, see equation (\ref{eq:themetric}). We will elaborate on this in subsequent sections.


\subsection{Seed functions for horizonless spacetimes}
\label{sec:seeds}

In the language of the solution generating technique outlined in Section \ref{sec:2}, we have given the first small step in the way to obtain a supersymmetric solution, that can be synthesized as

\begin{equation}
\label{eq:Hmulti}
H=\sum_a \frac{q_a}{r_a} \, , \hspace{1.5cm} \text{with} \hspace{0.5cm} q_a \in \mathbb{Z} \, , \hspace{0.5cm} \sum_a q_a = 1 \, .
\end{equation}

\noindent
The remaining seed functions in the Abelian sector $\Phi^i$, $L_i$ and $M$ are also harmonic,
\begin{equation}
\label{eq:Abmulti}
\Phi^i=k^i_0+\sum_a \frac{k^i_a}{r_a} \, , \hspace{1.5cm}
L_i=l^i_0+\sum_a \frac{l^i_a}{r_a} \, , \hspace{1.5cm}
M=m_0+\sum_a \frac{m_a}{r_a} \, ,
\end{equation}

\noindent
and from equation (\ref{eq:PhiIequation}) we readily obtain
\begin{equation}
\label{eq:breveA}
\breve{A}^i=  \sum_a k^{i}_a cos \theta_a d\psi_a \, .
\end{equation}

Notice that we imposed that the location of the singularities coincides with a Gibbons-Hawking center. With this requirement we will be able to avoid that the building blocks $h_I/f$ as defined in (\ref{eq:hIf}) become singular whenever any of the seed functions individually diverge. This is the mathematical version of what at the beginning of the section we called absence of brane sources, and it is the mechanism responsible of obtaining horizonless geometries\footnote{Clearly this naming is pointing at the physical origin of these potential singularities once the solutions are interpreted in the context of string theory.}. Also, the fact that the harmonic seed functions are singular at the Gibbons-Hawking centers is directly responsible for much of the very interesting physics captured by these solutions. Consequently, we would like the non-Abelian seed functions to display a similar qualitative behavior, i.e. $(\Phi^\alpha , L_\alpha) \vert_{r_a \rightarrow 0} \sim r_a^{-1}+\mathcal{O}(r_a^0)$.

Protogenov's $SU(2)$ colored monopole \cite{Protogenov:1977tq} is a solution to the Bogomol'nyi equation with this property, with only one single center. Colored monopoles are rather intriguing objects. They describe a point with unit local magnetic charge surrounded by a magnetic cloud that completely screens the charge as seen from infinity\footnote{The magnetic charge is defined as $p=\frac{\breve{g}}{4\pi}\int_{S^2} \frac{\Phi^\alpha \breve{F}^\alpha}{\sqrt{\Phi^\alpha \Phi^\alpha}}$.}. Despite its singular nature when interpreted in the context of Yang-Mills-Higgs theory, single center colored monopole solutions have been fruitfully used in the literature to obtain regular non-Abelian black holes in four- \cite{Meessen:2015nla,Hubscher:2008yz,Meessen:2008kb,Bueno:2014mea} and five-dimensional \cite{Meessen:2015enl,Ortin:2016bnl} theories of gauged supergravity. Their presence has an interesting impact on black hole thermodynamics, modifying the entropy without altering the mass.

Therefore, a family of well-suited non-Abelian seed functions $\Phi^\alpha$ is given by a multicenter generalization of colored monopoles, which we construct now. From now on we will assume the gauged group is $SU(2)$ for the sake of simplicity, so the index $\alpha$ can take three possible values. Nevertheless, following the ideas of Meessen and Ort\'in \cite{Meessen:2015nla}, it should be possible to embed these monopoles in a more general group $SU(N)$ and use them in the construction of smooth bubbling geometries in $SU(N)$-gauged supergravity.

Plugging in the Bogomoln'yi equation (\ref{eq:PhiIequation}) the ansatz of the \emph{hedgehog} form

\begin{equation}
\label{eq:multicolored}
\Phi^\alpha =- \frac{1}{\breve{g}P} \frac{\partial P}{\partial x^s} \delta_s^\alpha \, , \hspace{1.5cm}
\breve{A}^\alpha_\mu =- \frac{1}{\breve{g}P} \frac{\partial P}{\partial x^s} \varepsilon^\alpha\,_{\mu s} \, ,
\end{equation}

\noindent
we find that this configuration describes a monopole solution if $P$ is a harmonic function,
\begin{equation}
P=\lambda_0+\sum_a \frac{\lambda_a}{r_a} \, , \hspace{1.5cm} \lambda_0 \neq 0 \, .
\end{equation}

\noindent
Substituting back in (\ref{eq:multicolored}), we can write the solution as

\begin{equation}
\label{eq:multicoloredext}
\Phi^\alpha = \sum_a \frac{\lambda_a}{\breve{g} r_a^2 P} \delta_s^\alpha \frac{(x^s-x^s_a)}{r_a} \, , \hspace{1.5cm}
\breve{A}^\alpha_\mu = \sum_a \frac{\lambda_a}{\breve{g} r_a^2 P} \varepsilon^\alpha\,_{\mu s} \frac{(x^s-x^s_a)}{r_a} \, .
\end{equation}

\noindent
The Higgs field of the monopole is singular at the centers and vanishes at infinity
\begin{equation}
\lim_{r_a \rightarrow 0} \Phi^\alpha = \frac{k_a^\alpha}{ r_a} + \mathcal{O}(r_a^0) \, , \hspace{1cm}
\lim_{r \rightarrow \infty} \Phi^\alpha \sim \mathcal{O}(r^{-2}) \, , \hspace{1cm}
k_a^\alpha \equiv \delta_s^\alpha \frac{(x^s-x^s_a)}{\breve{g} r_a} \, .
\end{equation}

\noindent
This solution corresponds to a \emph{multicenter colored monopole} configuration. 

The last seed functions we need to find are $L_\alpha$, which are solutions of equation (\ref{eq:LIequation}), that we repeat here for convenience

\begin{equation}
\breve{\mathfrak{D}}^{2}L_{\alpha} 
-\breve{g}^{2} f_{\alpha\beta}{}^{\lambda}f_{\gamma\lambda}{}^{\rho}\Phi^{\beta}\Phi^{\gamma}L_{\rho} = 0\, .
\end{equation}

\noindent
We can solve this differential system by making use of the ansatz

\begin{equation}
\label{eq:Lmulti}
L_\alpha= - \frac{1}{\breve{g}P} \frac{\partial Q}{\partial x^s} \delta_s^\alpha \, , 
\end{equation}

\noindent
the equation reduces to the condition of $Q$ being harmonic. We choose $Q$ to be of the form
\begin{equation}
Q=\sum_a \frac{\sigma_a \lambda_a}{r_a} \, .
\end{equation}

\noindent
The functions $L_\alpha$ behave similarly to $\Phi^\alpha$ near the centers and at infinity

\begin{equation}
\lim_{r_a \rightarrow 0} L_\alpha = \frac{l_a^\alpha}{r_a} + \mathcal{O}(r_a^0) \, , \hspace{1cm}
\lim_{r \rightarrow \infty} L_\alpha \sim \mathcal{O}(r^{-2})  \, , \hspace{1cm}
l_a^\alpha \equiv \sigma_a \delta_s^\alpha \frac{(x^s-x^s_a)}{\breve{g} r_a}  \, ,
\end{equation}

\noindent
only differentiated by the presence of the parameters $\sigma_a$ in the near-center limit. The appearance of these factors will be fundamental for obtaining horizonless geometries.

After having fixed the general form of all the seed functions, we can start analyzing the regularity of the metric. In order to construct horizonless solutions we need to avoid having brane sources at the centers. In other words, we want the building blocks $h_I/f$ that constitute the metric function, given by (\ref{eq:hIf}), to remain finite at these points. Keeping the charges $q_a$ and $k^i_a$ arbitrary, it is possible to remove the brane sources by taking

\begin{equation}
\label{eq:lacoef}
l^I_a = -8 C_{IJK} \frac{k^J_a k^K_a}{q_a} \, .
\end{equation}

\noindent
Notice that this expression is valid in both the Abelian and the non-Abelian sector. In the former it fixes the value of the parameters $l^i_a$, while in the latter it fixes the parameters $\sigma_a$. Regularity of the metric at the centers also requires $\omega_5$ to be finite there, something that we achieve by choosing

\begin{equation}
\label{eq:macoef}
m_a = 8\sqrt{2} C_{IJK}  \frac{k^I_a k^J_a k^K_a}{q^2_a} \, .
\end{equation}

\noindent
These conditions, together with the fact that the Gibbons-Hawking metric looks locally like $\mathbb{R}^4$ near the centers (up to orbifold singularities), are enough to guarantee smoothness at these particular points.

The constant terms in the harmonic seed functions (\ref{eq:Abmulti}) define the solution at infinity. In order to have an asymptotically flat metric ($f_\infty \sim 1$, $\omega_{5,\infty} \sim 0$) we need to satisfy the constrains

\begin{equation}
k^i_0=0 \, , \hspace{1.5cm}
27 C^{ijk} l^i_0 l^j_0 l^k_0 = 1 \, , \hspace{1.5cm}
m_0 = -3\sqrt{2} \sum_{i,a} l_0^i k^i_a \, .
\end{equation}

\subsection{Closed timelike curves and bubbling equations}
\label{sec:CTC}

By using an ambipolar Gibbons-Hawking metric we are taking a clear risk: the spacetime metric might contain closed timelike curves (CTC's) or even be ill-defined at the critical surfaces where $H=0$. We now study the conditions under which CTC's are absent, so the microstate geometries are physically sensible. 

Let us expand the expression of the spacetime metric (\ref{eq:themetric}) and write it in the following manner
\begin{equation}
\label{eq:metricexp}
ds^2= f^2 dt^2 +2f^2 dt \omega-\frac{\mathcal{I}}{f^{-2} H^2} \left( d\varphi + \chi - \frac{\omega_5 H^2}{\mathcal{I}} \breve{\omega} \right)^2 - f^{-1} H \left( d\vec{x} \cdot d\vec{x}-\frac{\breve{\omega}^2}{\mathcal{I}} \right) \, ,
\end{equation}

\noindent
where $\mathcal{I}$ is defined as

\begin{equation}
\mathcal{I} \equiv f^{-3} H - \omega_5^2 H^2 \, .
\end{equation}

There is one general restriction that needs to be satisfied in order to avoid the presence of CTC's

\begin{equation}
\label{eq:CTC1}
\mathcal{I} \geq 0 \, .
\end{equation}

\noindent
Apparently there is one additional condition, $f^{-1} H \geq 0$, but this is implied by the inequality in (\ref{eq:CTC1}). Let us express this condition in more detail by evaluating $\mathcal{I}$ in terms of the seed functions
\begin{equation}
\label{eq:CTC}
\begin{array}{rcl}
\mathcal{I} &=& -M^2 H^2- 18 \left( \Phi^I L_ I \right)^2 -32\sqrt{2} M C_{IJK} \Phi^I \Phi^J \Phi^K - 6\sqrt{2} M H L_ I \Phi^I \\ \\
& & +27 H C^{IJK} L_I L_J L_K + 3^4 2^3 C^{IJK} C_{KLM} L_I L_J \Phi^L \Phi^M \geq 0 \, .
\end{array}
\end{equation}

The first point to notice is that the form of this expression coincides with that of ungauged supergravity originally derived in \cite{Bena:2005va}, where it was identified as the quartic invariant of $E_{7(7)}$. The analysis of the positivity of this quantity is hard to do in general, although we can assert that this bound can be satisfied for large families of configurations. The reason behind this statement is that this has been shown to be the case for ungauged supergravities, and many techniques to construct solutions satisfying this bound have been developed. In any case, it is fair to say that this restriction definitely makes the process of constructing explicit solutions more complicated.

There is one additional factor that can result in the appearance of CTC's, and this is the formation of Dirac-Misner strings. Those arise when the integrability condition of the last differential equation that still remains to be solved, (\ref{eq:omegaequation2}), is not satisfied. This condition is obtained acting with the operator $d \star_3$ in that expression, which gives

\begin{equation}
\left\{
H \nabla^2 M-M\nabla^2 H
+3\sqrt{2} ( \Phi^{i} \nabla^2L_{i}
-L_{i}\nabla^2 \Phi^{i} + \Phi^{\alpha} \breve{\mathfrak{D}}^2L_{\alpha}- L_{\alpha}\breve{\mathfrak{D}}^2 \Phi^{\alpha}  )
\right\}=0 \, .
\end{equation}

\noindent
This condition is identically satisfied as a consequence of equations (\ref{eq:Hequation})-(\ref{eq:LIequation}) everywhere except at the centers, where technically those equations cease to apply. The bubbling equations are algebraic constrains that guarantee that the integrability condition is satisfied everywhere, setting the requirements that avoid the presence of Dirac-Misner strings. 

To make further progress it is convenient to define the symplectic vector of seed functions

\begin{equation}
S^M=\left( H, 3\sqrt{2} \Phi^I , M , L_I \right) \, , \hspace{2cm} S_M=\left( M , L_I, -H, - 3\sqrt{2} \Phi^I  \right) \, ,
\end{equation}
and a symplectic vector of charges at each center

\begin{equation}
Q^M_a=\left( q_a, 3\sqrt{2} k^I_a , m_a, l^I_a\right) \, , \hspace{2cm} Q_{M,a}=\left( m_a , l^I_a, -q_a, - 3\sqrt{2} k_a^I  \right) \, .
\end{equation}

\noindent
Now we can write the integrability condition as

\begin{equation}
S^M \breve{\mathfrak{D}}^2 S_M= 0 \, . 
\end{equation}

Interestingly the non-Abelian sector vanishes in the last expression due to the symplectic product and the expression is reduced to $S^m Q_{m,a} \delta (\vec{x}-\vec{x}_a)= 0$ with the understanding that $S^m$ and $Q^m_a$ are the components of the symplectic vectors in the Abelian sector. Then, one could naively expect that the bubbling equations coincide with those in the case of ungauged supergravity theories. However, this does not happen because the charges $l^i_a$ are affected by the presence of the non-Abelian fields according to (\ref{eq:lacoef}). After a few lines of algebraic computation, the resulting bubbling equations are conveniently written as

\begin{equation}
\label{eq:bubbling}
\sum_{b \neq a} \frac{q_a q_b}{r_{ab}} \left[  C_{ijk} \Pi^i_{ab} \Pi^j_{ab} \Pi^k_{ab} -\frac{1}{2\breve{g}^2} \Pi^0_{ab}  \mathbb{T}_{ab} \right]  = 
\frac{3}{8} l_0^i \left( \sum_{b}  q_a k^i_b - k^i_a \right) \, ,
\end{equation}

\noindent
where $\Pi^i_{ab}$ is the $i^{\text{th}}$- flux threading the 2-cycle $\Delta_{ab}$ and $\mathbb{T}_{ab}$ contains information about the topological charge associated to the centers $a$ and $b$, see (\ref{eq:topocenter}) 

\begin{equation}
\Pi^i_{ab} \equiv \left( \frac{k^i_b}{q_b}-\frac{k^i_a}{q_a} \right) \, , \hspace{2cm}
\mathbb{T}_{ab} \equiv \breve{g}^2 \left( \frac{k^\alpha_a k^\alpha_a}{q_a^2} + \frac{k^\alpha_b k^\alpha_b}{q_b^2} \right) \, .
\end{equation}

We are now ready to integrate (\ref{eq:omegaequation2}). It is convenient to decompose the 1-form $\breve{\omega}$ into two parts, $\breve{\omega}^A$ and $\breve{\omega}^B$, satisfying
\begin{eqnarray}
\label{eq:omegaA}
\star_{3} d \breve{\omega}^A &=&  H dM-MdH +3\sqrt{2} ( \Phi^{i} d L_{i} -L_{i} d \Phi^{i} ) \, , \\
\label{eq:omegaB}
\star_{3} d \breve{\omega}^B &=& 3\sqrt{2} ( \Phi^{\alpha} \breve{\mathfrak{D}}L_{\alpha} -L_{\alpha}\breve{\mathfrak{D}} \Phi^{\alpha} ) \, ,
\end{eqnarray}

\noindent
The first equation can be solved independently for each pair of centers $(a,b)$, with $\breve{\omega}^A=\sum_a \sum_{b > a} \breve{\omega}^A_{ab}$. For each pair we use adapted coordinates such that $\vec{x}_a=(0,0,0)$ and $\vec{x}_b=(0,0,-r_{ab})$, with spherical angles given by 

\begin{equation}
x^1_{ab}= r_a sin \theta_{ab} sin \psi_{ab} \, \hspace{1cm}
x^2_{ab}= r_a sin \theta_{ab} cos \psi_{ab} \, \hspace{1cm}
x^3_{ab}= - r_a cos \theta_{ab}  \, .
\end{equation}

\noindent
Upon substitution of the seed functions $H,M,L_{i},\Phi^{i}$, (\ref{eq:omegaA}) can be written as
\begin{equation}
  \begin{array}{rcl}
\star_{3} d \breve{\omega}^A_{ab} 
& = &
\dfrac{Q_{m,a} Q^m_b }{r_{ab}} 
\bigg\{ 
{\displaystyle
-\frac{1}{r_a^{2}}\left[1-\frac{r_{ab} +
    r_a}{r_{b}}+\frac{r_a r_{ab} \left( r_a+r_{ab}
    \right)}{r_{b}^{3}} \left(1-\cos{\theta_{ab}} \right) \right] dr_a 
}
\\
& & \\
& & 
+ 
{\displaystyle
\left[ \frac{r_{ab} \sin{\theta_{ab}}}{r_{b}^{3}} \left(r_a- r_{ab} \right) \right] d\theta_{ab} 
}
\bigg\} \, ,
\\
\end{array}
\end{equation}

\noindent
being $r_b$ the radial distance as measured from $\vec{x}_b$. A solution can be readily found provided $\breve{\omega}^A_{ab}$ has only one non-vanishing component, $\breve{\omega}^A_{ab,\psi_{ab}}$

\begin{equation}
\label{eq:omegaAsol}
\breve{\omega}^{A}_{ab}
=
\frac{8\sqrt{2} q_a q_b}{r_{ab}} \left[  C_{ijk} \Pi^i_{ab} \Pi^j_{ab} \Pi^k_{ab} -\frac{1}{2\breve{g}^2} \Pi^0_{ab}  \mathbb{T}_{ab} \right]
\left(\cos{\theta_{ab}}-1\right)
\left(1-\frac{r_a+ r_{ab}  }{r_{b}} \right) d\psi_{ab} \, .
\end{equation}

Now we turn our attention to (\ref{eq:omegaB}). Notice that this expression contains three-point interactions due to the presence of the connection $\breve{A}^\alpha$ in the covariant derivative, so at first sight its structure is more involved than that of its Abelian counterpart. However, despite this complexity, the general solution for an arbitrary number of centers can be found. It is most remarkable that the interactions among all of them can be written in a very compact form! We obtain

\begin{equation}
\label{eq:omegaBsol}
\breve{\omega}^B = \frac{3 \sqrt{2} \varepsilon_{r s t}}{\breve{g}^2 P^2} \frac{\partial Q}{\partial x^s} \frac{\partial P}{\partial x^t} dx^r \, .
\end{equation}

While deriving (\ref{eq:omegaAsol}) and (\ref{eq:omegaBsol}) we have assumed that the integrability condition is satisfied by making use of the bubbling equations (\ref{eq:bubbling}). As a consistency check we can perform an inspection to confirm the absence Dirac-Misner strings in $\breve{\omega}^A$ and $\breve{\omega}^B$. For the former, it is straightforward to verify that the only component of the one form, $\breve{\omega}^A_{ab,\psi_{ab}}$, vanishes when the coordinate $\psi_{ab}$ is not well defined. In particular this happens along the $x^3_{ab}$ axis both in the positive direction, where $\left(1-\frac{r_a+ r_{ab}  }{r_{b}} \right)\vert_{x^{3,+}_{ab}}=0$, and in the negative direction, with $\left(\cos{\theta_{ab}}-1\right)\vert_{x^{3,-}_{ab}}=0$. In the case of the latter it suffices to check that $\breve{\omega}^B$ is regular at the centers as a consequence of the antisymmetric character of the 1-form components.



\subsection{Fluxes and topological charge}
\label{sec:flux}

We now turn our attention to the vector fields. We shall recall their expressions
\begin{eqnarray}
\label{eq:vec2}
A^I &=&-\sqrt{3} h^I f (dt+\omega) + \hat{A}^I \, ,  \\ \label{eq:Kronh2}
\hat{A}^I &=& -2\sqrt{6} \left[ -\Phi^I H^{-1} \left( d\varphi+\chi \right) + \breve{A}^I \right] \, , 
\end{eqnarray}

\noindent 
where $\breve{A}^I$ is determined in terms of $\Phi^I$ by the Bogomol'nyi equation (\ref{eq:PhiIequation}) and whose explicit form is (\ref{eq:breveA}) in the Abelian sector and (\ref{eq:multicolored}) in the non-Abelian. From these expressions we see that these fields can be understood in terms of three layers: the physical vectors $A^I$, a four-dimensional instanton $\hat{A}^I$ with selfdual field strength and a three-dimensional static magnetic monopole $\breve{A}^I$. Each of them is used to build up those preceding it, in a configuration that resembles the structure of the Russian \emph{matryoshka} dolls.

In the Abelian sector $\breve{A}^i$ describes a configuration with several Dirac monopoles, which is singular due to the presence of Dirac strings attached to each center. These strings are eliminated in $\hat{A}^i$ by the new term in (\ref{eq:Kronh2}), although this term introduces new strings in the compact direction $\varphi$,
\begin{equation}
\lim_{r_a \rightarrow 0} \hat{A}^i \sim -2\sqrt{6}\left[ - \frac{k^i_a}{q_a} \left( d\varphi + q_a cos\theta_a d\psi_a \right) + k^i_a cos\theta_a d\psi_a \right] \sim 2\sqrt{6} \frac{k^i_a}{q_a} d\varphi  \, .
\end{equation}

\noindent
The component in the local coordinate $\psi_a$ is compensated by the new term, but now $\hat{A}^i_\varphi$ is finite at the centers, where the coordinate $\varphi$ is not well defined. Besides $\hat{A}^i$ is not regular either at the critical surfaces characterized by $H=0$. Yet again, this singularity is cured at the next stage and the physical vectors $A^i$ are globally regular up to gauge transformations. In this case the first term in (\ref{eq:vec2}) compensates the divergence at the critical surface,

\begin{equation}
\lim_{H \rightarrow 0} \left(-\sqrt{3} h^i f \omega_5 (d\varphi + \chi )  \right) = -2\sqrt{6} H^{-1} \Phi^i (d\varphi + \chi ) + \mathcal{O} (H^0) \, ,
\end{equation}

\noindent 
without introducing any anomaly elsewhere, which is guaranteed because $\omega$ has been designed to be free of Dirac-Misner strings.

To every non-trivial 2-cycle at the ambipolar space it is naturally associated a magnetic flux for each vector, defined as the integral of the field strength $ F^i$ along the 2-cycle. To compute this quantity we make use of our standard decomposition for $A^i$, which is valid everywhere except at the centers. Nevertheless since the field strength is globally regular the flux can be equally computed by taking the integral along the 2-cycle with the poles excised. In this region the integrand is an exact form and we can make use of Stokes' theorem. We get

\begin{equation}
\Pi^i_{ab} \equiv \frac{1}{(2\sqrt{6}) 4\pi} \int_{\Delta_{ab}} \, F^i = \left( \frac{k^i_b}{q_b}-\frac{k^i_a}{q_a} \right) \, .
\end{equation}

We now consider the non-Abelian sector. Our recipe for constructing solutions of $\mathcal{N}=1$, $d=5$ SEYM theory naturally incorporates Kronheimer's scheme \cite{KronMT}, that relates any static monopole $\breve{A}^\alpha$ to an instanton over a Gibbons-Hawking base, $\hat{A}^\alpha$, through equation (\ref{eq:Kronh2}). For example, in \cite{Bueno:2015wva} this mechanism has been utilized to \emph{oxidize} the single center colored monopole, that has turned out to be the counterpart of the BPST instanton \cite{BELAVIN197585}. On the other hand, Etesi and Hausel showed in \cite{Etesi:2002cc} that families of regular Yang-Mills instantons over an Asymptotically Locally Euclidean space (ALE) are related to multicenter colored monopoles in Kronheimer's scheme\footnote{In fact, to the best of our knowledge, multicenter colored monopoles have only appeared in the literature so far in \cite{Etesi:2002cc}, where they are used as valuable intermediates for computing the topological charge of their instanton counterparts.}. However, although our instanton is related to the same monopole, it is necessarily different than the Etesi-Hausel solution because they are defined on different bases: our Gibbons-Hawking space is ambipolar, not ALE. In particular this means that our instanton is singular at the critical surfaces. This is cured for the five-dimensional physical vector in the same manner than it is for the Abelian vectors.

Even though the instanton $\hat{A}^\alpha$ is ill-defined at the critical surfaces, we would like to study if we can associate to it a topological charge, also known as instanton number\footnote{It would be very interesting to study rigorously the construction of $SU(2)$ fiber bundles over ambipolar Gibbons-Hawking bases, but this goes beyond the scope of the present work.}. As we are about to see this quantity is finite even though the connection blows up. We define the topological charge as

\begin{equation}
\label{eq:topodef}
\mathbb{T}= \frac{g^2}{32\pi^2} \int_{\mathcal{M}_4 \backslash S} d^4\Sigma  \hat{F}^2 \, ,
\end{equation}
where $d^4\Sigma$ is the volume form of the manifold, $\hat{F}^2$ is the scalar obtained by taking the trace of the field strength contracted with itself, $\hat{F}^2 \equiv \hat{F}^\alpha_{\mu\nu} \hat{F}^{\alpha \, \mu\nu}$, and $\mathcal{M}_4\backslash S$ is the ambipolar space without the critical surfaces. These have to be necessarily removed because the canonical volume form associated to the metric vanishes there and the above integral cannot be defined over them. To perform the calculation it is convenient to work in the following flat frame of the cotangent bundle
\begin{equation}
e^0=s \vert H \vert^{-1/2} (d\varphi+\chi) \, , \hspace{1cm} e^a= \vert H \vert^{1/2} dx^s \delta_s^a \, , \hspace{1cm} \epsilon^{0123}=\epsilon_{0123}=1 \, .
\end{equation}

\noindent
where $s$ is $+1$ when $H$ is positive and $-1$ when $H$ is negative. The volume form is expressed in terms of the vielbeins as $e^0 \wedge e^1 \wedge e^2 \wedge e^3 = H d\varphi \wedge d^3x$, where $d^3x$ is a shorthand for $ dx^1 \wedge dx^2 \wedge dx^3$. The gauge field strength is obtained from (\ref{eq:Kronh2}) and its components in this coframe are

\begin{equation}
\label{eq:Fframe}
\hat{F}^\alpha_{0a}=-2\sqrt{6}s \breve{\mathfrak{D}}_a \left( \Phi^\alpha H^{-1} \right) \, , \hspace{1cm}
\hat{F}^\alpha_{ab}= -2\sqrt{6}s \left[ H^{-1} \breve{F}^\alpha_{ab} - H^{-2} \Phi^\alpha (d\chi)_{ab} \right] \, .
\end{equation}

\noindent
Substituting back into (\ref{eq:topodef}), using (\ref{eq:Hequation}), (\ref{eq:PhiIequation}) and integrating by parts we get

\begin{equation}
\mathbb{T}= \frac{\breve{g}^2}{32\pi^2} \int_{\mathcal{M}_4 \backslash (S \cup X)} d\varphi \wedge d^3x \left[ 2 \nabla^2 \left( \frac{\Phi^\alpha \Phi^\alpha}{H} \right) -4  H^{-1} \Phi^\alpha\breve{\mathfrak{D}}^2 \Phi^\alpha + 2  H^{-2} \Phi^\alpha \Phi^\alpha \nabla^2 H \right]   \, .
\end{equation}

\noindent
Notice that in this step the centers have also been removed from the integration space because the decomposition (\ref{eq:Kronh2}) is not well-defined there. This does not change the value of the integral because $\hat{F}^2$ is regular at these points. The second and third terms in the above expression vanish identically in the region. We can integrate on $\varphi$ and apply Stokes theorem to get

\begin{equation}
\label{eq:topo2}
\mathbb{T}=\frac{\breve{g}^2}{4\pi} \int_{V^3} d^3x \nabla^2 \left( \frac{\Phi^\alpha \Phi^\alpha}{H} \right) = \frac{\breve{g}^2}{4\pi} \int_{\partial V^3} d^2 \Sigma \, n_{a} \partial_a \left( \frac{\Phi^\alpha \Phi^\alpha}{H} \right)  \, .
\end{equation}

\noindent
Here $V^3$ is $\mathbb{R}^3$ with the centers and the critical surfaces excised, $d^2 \Sigma$ is the volume form induced on $\partial V^3$ and $n_{a}$ are the components of a unit vector normal to $\partial V^3$. Thus the problem is reduced to a computation at the boundary of $V^3$, which is composed of the critical surfaces, the centers and infinity. Formally at the critical surfaces we receive an infinite contribution to the topological charge, but notice that each connected critical surface is the boundary of two disconnected regions of $V_3$ and therefore it appears twice in the computation. Since the normal unitary vector $\vec{n}$ has opposite direction in each case, both infinite contributions cancel out because $\lim_{\vec{x}\rightarrow \partial V^3}\partial_a \left( \frac{\Phi^\alpha \Phi^\alpha}{H} \right) \vert n_a \vert$ takes the same value when $\vec{x}$ is evaluated at both sides of the critical surface. 

After having got rid of the critical surfaces, the computation of (\ref{eq:topo2}) is straightforward. The contributions at each center and at infinity are

\begin{equation}
\label{eq:topocenter}
\mathbb{T}_a = \breve{g}^2  \frac{k^\alpha_a k^\alpha_a}{q_a} \, , \hspace{1.5cm}
\mathbb{T}_\infty = 0 \, ,
\end{equation}

\noindent
Assuming that we placed non-Abelian seed functions at every center, the total topological charge is

\begin{equation}
\mathbb{T} = \sum_a \frac{1}{q_a}  \, .
\end{equation}

\subsection{Critical surfaces}
\label{sec:critical}

As we have already discussed at previous stages, the critical surfaces defined by having $H=0$ are worth special attention. Not only is the ambipolar Gibbons-Hawking metric ill-defined there, but also many of the other auxiliary building blocks that make up the solution contain inverse powers of $H$. Nevertheless, the spacetime metric and all physical fields remain completely regular at the critical surfaces. It is interesting to illustrate in some detail how this happens.

Let us consider the metric as written in (\ref{eq:metricexp}). In the purely spatial part there are no singularities in these surfaces because the product $f^{-1} H$ defines a finite positive quantity,

\begin{equation}
\lim_{H \rightarrow 0} f^{-1} H = 8 \left( C_{IJK} \Phi^I \Phi^J \Phi^K \right)^{2/3} + \mathcal{O}(H) \, ,
\end{equation}
and $\mathcal{I}$ is also regular, as easily seen from its expression in terms of the seed functions (\ref{eq:CTC}). Of course, this is only possible because $\lim_{H \rightarrow 0} f \sim 0$ and this, in particular, means that the critical surfaces are determined by the vanishing of the norm of the Killing vector that generates time translations, $V=\partial_t$, $V^\mu V_\mu = f^2$. 

One might get worried by this statement, since timelike supersymmetric solutions in supergravity quite frequently have event horizons at the regions where the timelike Killing vector becomes null. Happily this does not happen here. First, because as we just saw the spatial part remains regular, and second, because of the presence of the additional finite term in the metric that keeps the determinant non-vanishing at these regions,
\begin{equation}
\lim_{H \rightarrow 0} f^2 \omega_5 dt (d\varphi+\chi)= \frac{1}{2\sqrt{2}} \left( C_{IJK} \Phi^I \Phi^J \Phi^K \right)^{-1/3} dt (d\varphi+\chi) + \mathcal{O}(H)  \, .
\end{equation}

\noindent
Then any massive particle sitting at the surface is unavoidably dragged along some spatial direction. Critical surfaces have the same properties as the boundary of an ergosphere, except from the fact that they do not actually surround an ergosphere since the Killing vector $V$ remains timelike at both of their sides. As a consequence of this they have been named \emph{evanescent ergosurfaces} \cite{Gibbons:2013tqa}.

In the previous subsection we already showed that the physical vectors are well-behaved at the evanescent ergospheres. The physical scalars, constructed by $\phi^x \equiv h_I/h_0$, are also regular here
\begin{equation}
\lim_{H \rightarrow 0} \phi^x=\frac{C_{xIJ}\Phi^I \Phi^J }{ C_{0LM} \Phi^L \Phi^M} + \mathcal{O}(H)  \, .
\end{equation}

%

\section{Final comments}
\label{sec:final}

The set of continuous parameters $\lambda_a$ that appear in the definition of the colored monopole, (\ref{eq:multicolored}), have no impact on the physics of the solution neither at the centers nor at infinity, but they do affect the physical fields at intermediate regions. This means that the geometry of a particular solution can be continuously distorted in some manner as long as the modification does not introduce CTC's. Therefore we can build a \emph{classically} infinite number of microstate geometries with the same topology for the same black hole or black ring. 

It is useful to explain in some detail why these parameters are special in this sense. First, one has to notice that asymptotically the non-Abelian seed functions $\Phi^\alpha$ are subleading with respect to the Abelian seed functions $\Phi^i$ (\ref{eq:Abmulti}). Second, the functions $\Phi^\alpha$ have the same limit at leading order at all the centers, whose value is independent of these parameters. These characteristics imply that the mass, angular momenta and electric charges of the solution are invisible to the parameters $\lambda_a$. The size of the bubbles are also unaffected by them, see (\ref{eq:bubbling}).

The \emph{colored} non-Abelian black hole solutions discovered so far are constructed from a single-center colored monopole. They incorporate one parameter, say $\lambda_1$, interpreted as the \emph{size} of the instanton field of the solution, that modifies the geometry outside the horizon but does not alter any of the \emph{observables} of the solution, like the mass, entropy, electric charges or instanton number. In this context this parameter is interpreted as non-Abelian hair. On the other hand microstate geometries have one parameter for each center. Although we do not have a complete interpretation of the multicenter instanton field contained in these solutions, preliminary analysis based on the expansion of the instanton field $\hat{A}^\alpha$ near the centers suggest that each parameter codifies the information of the size of an instanton placed at the corresponding center whose individual topological charge is $1/q_a$.

\begin{figure}[b!]
\centering
\includegraphics[width=1\textwidth]{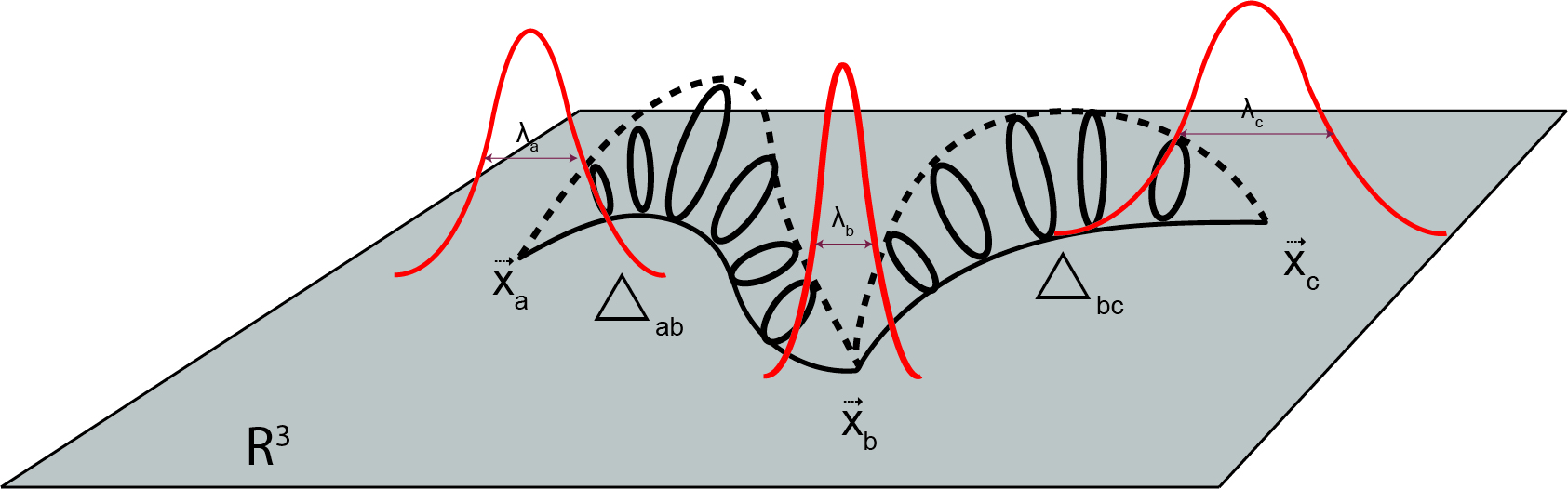}
\caption{\small{Representation of the multicenter instanton on the Gibbons-Hawking space.}}
\label{figure}
\end{figure}

On the other hand, the gauge coupling constant $\breve{g}$ controls the relative weight of the non-Abelian versus the Abelian fields. The closer this parameter is to zero the more influent the non-Abelian ingredients are. This is in particular reflected in the bubbling equations (\ref{eq:bubbling}), from what we see that the size of the bubble can be dominated by one or the other contributions for different values of the coupling constant.

 Clearly these solutions require further study. The explicit construction of concrete solutions with specific charges would be of course very interesting. Work in this direction is in progress \cite{inprogress}.

\section*{Acknowledgments}

This work would not have been possible without the useful advice of Iosif Bena, Pablo Bueno, Tom\'as Ort\'in, Carlos S. Shahbazi, David Turton and Nick Warner.
The project has been supported in part by the Spanish Ministry of Science and Education grant FPA2012-35043-C02-01, the Centro de Excelencia Severo Ochoa Program grant SEV-2012-0249, the \textit{Severo Ochoa} pre-doctoral grant SVP-2013-067903, the EEBB-I-16-11563 grant and the ERC Starting Grant 259133 \emph{ObservableString}. Last but not least, I am thankful to Nieves L\'opez for drawing Figure \ref{figure} and for asking me to marry her.

\appendix

\section{The theory}
\label{sec:theory}

In this appendix we give a very brief, workable description of SEYM theories and their known analytic solutions adapted to the purpose of this letter. $\mathcal{N}=1,d=5$ gauged supergravities can be interpreted as the minimal supersymmetric realization of Einstein-Yang-Mills-Higgs theories\footnote{Those were first considered in \cite{GUNAYDIN1985573}, see \cite{Meessen:2015enl,Bellorin:2006yr,Bellorin:2007yp,Bergshoeff:2004kh} for more detailed expositions in our same conventions.}. They describe the coupling between a supergravity multiplet and $n_v$ vector multiplets, a subset of which transform under the local action of a non-Abelian group. The supergravity multiplet is constituted by the graviton $e^{a}{}_{\mu},$ the gravitino $\psi_{\mu}^{i}$ and the graviphoton $A^0_{\mu}$, while each vector multiplet, labeled by $x=1,....,n_{v}$, contains a real vector field $A^{x}{}_{\mu},$ a real scalar $\phi^{x}$ and a gaugino $\lambda^{i\, x}$. The vector fields can be collectively denoted as $A^{I}{}_{\mu}$, with $\{ I,J,\ldots=0,1,\cdots,n_{v} \}$. The set over which these indices take values is conveniently split in two sectors denoted as $\{ i,j,\cdots = 0, \cdots, i_{max} \}$ and $\{ \alpha, \beta, \cdots= i_{max}+1, \cdots, n_v \}$, referred as the Abelian and the non-Abelian sectors respectively.

The $n_v$ scalars $\phi^x$ parametrize a $\sigma$-model equipped with a Riemannian metric $g_{xy}$ and can be understood as coordinates on a scalar manifold. On general grounds the $\sigma$-model metric is invariant under coordinate transformations in the scalar manifold of the form 
\begin{equation}
\label{eq:scalartransf}
\delta_{\Lambda} \phi^x=-\hat{g} c^I k_{I}{}^{x} \, ,
\end{equation}
where $\hat{g}$ is interpreted as the gauge coupling constant (see below) and $k_{I}{}^{x}(\phi)$ is a set of Killing vectors of the scalar metric\footnote{Here the index $I$ is for labeling each one of these vectors. We use it in order to keep notation simple, and it should be understood that the Killing vectors will be non-zero only for a subset of the possible values of the index.}. The requirement that the $\sigma$-model is compatible with the supersymmetric structure that controls the coupling between scalars and vectors gives rise to the mathematical construct known as Real Special Geometry, see \cite{Bergshoeff:2004kh,Ortin:2015hya}, that completely characterizes the supergravity theory. Then, a Killing vector of the scalar metric generates an isometry of the full supergravity theory if it respects the real special structure of the theory, see Appendix H in \cite{Ortin:2015hya}.

The parameters that generate these isometries in the non-Abelian sector are spacetime functions, i.e. $c^\alpha=c^\alpha(x)$, while the corresponding Killing vectors satisfy the algebra
\begin{equation}
\left[ k_\alpha , k_\beta \right] = - f_{\alpha \beta} \, ^\gamma k_\gamma \, ,
\end{equation}
 where $f_{\alpha\beta}\,^{\gamma}$ are the structure constants of some non-Abelian group (we will often use the notation $f_{IJ}\,^{K}$, understanding that the structure constants just vanish whenever any index take values in the Abelian sector). 

The vectors in the non-Abelian sector, i.e. $A^\alpha \,_\mu$, play the role of gauge fields under the action of (\ref{eq:scalartransf}). That is, they transform in an appropriate way such that the covariant derivative of the scalars defined as

\begin{equation}
\mathfrak{D}_{\mu}\phi^{x}= \partial_{\mu}\phi^{x}+\hat{g}A^{\alpha}{}_{\mu}k_{\alpha}{}^{x} \, ,
\end{equation}
transforms, indeed, covariantly. The field strengths are defined in the standard manner in both the Abelian and non-Abelian sectors,
\begin{equation}
F^{I}{}_{\mu\nu}=2\partial_{[\mu}A^{I}{}_{\nu]}+\hat{g}f_{JK}{}^{I}A^{J}{}_{\mu}A^{K}{}_{\nu} \, .
\end{equation}

We will set all the fermionic fields to zero, which is always a consistent truncation in these theories. The bosonic action of $\mathcal{N}=1,d=5$ SEYM is given by

\begin{equation}
\label{eq:action}
\begin{array}{rcl}
S & = &  {\displaystyle\int} d^{5}x\sqrt{g}\
\biggl\{
R
+{\textstyle\frac{1}{2}}g_{xy}\mathfrak{D}_{\mu}\phi^{x}
\mathfrak{D}^{\mu}\phi^{y}
-{\textstyle\frac{1}{4}} a_{IJ} F^{I\, \mu\nu}F^{J}{}_{\mu\nu}
+\tfrac{1}{12\sqrt{3}}C_{IJK}
{\displaystyle\frac{\varepsilon^{\mu\nu\rho\sigma\lambda}}{\sqrt{g}}}
\left[
F^{I}{}_{\mu\nu}F^{J}{}_{\rho\sigma}A^{K}{}_{\lambda}
\right.
\\ \\ & & 
\left.
-\tfrac{1}{2}\hat{g}f_{LM}{}^{I} F^{J}{}_{\mu\nu} 
A^{K}{}_{\rho} A^{L}{}_{\sigma} A^{M}{}_{\lambda}
+\tfrac{1}{10} \hat{g}^{2} f_{LM}{}^{I} f_{NP}{}^{J} 
A^{K}{}_{\mu} A^{L}{}_{\nu} A^{M}{}_{\rho} A^{N}{}_{\sigma} A^{P}{}_{\lambda}
\right]
\biggr\}.
\end{array}
\end{equation}

The Real Special Geometry, and therefore the full supergravity theory, is completely determined by the constant symmetric tensor $C_{IJK}$. In particular the $\sigma$-model metric $g_{xy}(\phi)$ and the kinetic matrix $a_{IJ}(\phi)$ are directly derived from this tensor, see for example \cite{Meessen:2015enl} for the explicit expressions.

We make use of the SU$(2)$-gauged ST$[2,6]$ model, that contains $n_v=5$ vector multiplets and the constant symmetric tensor $C_{IJK}$ that characterizes it has the following non-vanishing components
\begin{equation}
C_{0xy}= \tfrac{1}{6}\eta_{xy}\, ,
\mbox{where}
\,\,\,\,\,
(\eta_{xy}) = \mathrm{diag}(+-\dotsm -)\, ,
\,\,\,\,\,
\mbox{and}
\,\,\,\,\,
x,y=1,\cdots,5\, .
\end{equation}

\section{Procedure for constructing solutions}
\label{sec:recipe}

\begin{enumerate}
\item Timelike supersymmetric solutions of $\mathcal{N}=1$, $d=5$ SEYM with a spacelike isometry are constructed from a set of $(2n_v+4)$ seed functions defined on $\mathbb{E}^3$. These are denoted\footnote{Notice that the seed functions $\Phi^I$ should not be confused with the physical scalars $\phi^x$ appearing in the action (\ref{eq:action}).} as $M,H,\Phi^{I},L_{I}$ and satisfy the following equations

\begin{eqnarray}
\label{eq:Mequation}
d\star_{3} d M 
& = &
0\, ,
\\
& & \nonumber \\
\star_{3}dH -d\chi  
& = &
0\, ,
\\
& & \nonumber \\ 
\star_{3}\breve{\mathfrak{D}} \Phi^{I}
- \breve{F}^{I}
& = &
0\, ,
\\
& & \nonumber \\
\breve{\mathfrak{D}}^{2}L_{I} 
-\breve{g}^{2} f_{IJ}{}^{L}f_{KL}{}^{M}\Phi^{J}\Phi^{K}L_{M}
& = &
0\, ,  
\\
& & \nonumber \\
\label{eq:omegaequationB}
\star_{3} d \breve{\omega}
-
\left\{
H dM-MdH
+3\sqrt{2} ( \Phi^{I} \breve{\mathfrak{D}}L_{I}
-L_{I}\breve{\mathfrak{D}} \Phi^{I} )
\right\}
& =&
0\, ,
\end{eqnarray}

for some 1-forms $\chi, \breve{\omega}$ and $\breve{A}^{I}$ (with field strength $\breve{F}^I$) defined also in $\mathbb{E}^{3}$. Here the covariant derivative $\breve{\mathfrak{D}}$ is defined in three-dimensional Euclidean space with respect to the gauge field $\breve{A}^I$ for objects transforming in the (dual) adjoint representation. More explicitly,

\begin{equation}
\breve{\mathfrak{D}} \Phi^I = d \Phi^I + \breve{g} f_{JK}\,^{I} \breve{A}^J \Phi^K \, , \hspace{2cm} \breve{\mathfrak{D}} L_I = d L_I + \breve{g} f_{IJ}\,^{K} \breve{A}^J L_K \, .
\end{equation}

Two subtleties about these expressions are worth mentioning. First, notice that the structure constants are only non-trivial in the non-Abelian sector so the covariant derivative reduce to the standard exterior derivative in the Abelian sector. Second, the gauge coupling constant in this expression is rescaled with respect to the \emph{physical} gauge constant appearing in the action of the theory\footnote{This fact is an indirect consequence of the rescaling factor appearing in equation (\ref{eq:instantondec2}).}, $\hat{g}=-{\breve{g}}/{2\sqrt{6}}$.

\item  Using the seed functions, the five-dimensional fields of the solution are obtained as follows:

\begin{enumerate}
\item We define the intermediate building blocks
\begin{equation}
\label{eq:hIf2}
h_{I}/f 
= 
L_{I}+8C_{IJK}\Phi^{J}\Phi^{K}/H\, ,
\end{equation}
that can be used to compute the physical scalars
\begin{equation}
\label{eq:scalars2}
\phi^{x} \equiv h_{x}/h_{0}\, ,
\end{equation}
and the metric function
\begin{equation}
\label{eq:f3symmetric2}
  \begin{array}{rcl}
f^{-3}
& = &
3^{3} C^{IJK}L_{I}L_{J}L_{K}
+3^{4}\cdot 2^{3}  C^{IJK}C_{KLM}L_{I}L_{J}\Phi^{L}\Phi^{M}/H
\\
& & \\
& &
+3\cdot 2^{6}L_{I}\Phi^{I}C_{JKL}\Phi^{J}\Phi^{K}\Phi^{L}/H^{2}
+2^{9}\left(C_{IJK}\Phi^{I}\Phi^{J}\Phi^{K}\right)^{2}/H^{3}\, .
\end{array}
\end{equation}
This is derived from the Real Special Geometry constrain $27 C^{IJK} h_I h_J h_K = 1$, which is valid for symmetric scalar manifolds\footnote{This is always the case in the supergravity models that we consider here. In this expression, $C^{IJK}\equiv  C_{IJK}$.}. In these spaces we can also define
\begin{equation}
h^I=27 C^{IJK} h_J h_K \, .
\end{equation}

\item The spacetime metric is of the conformastationary form
\begin{equation}
\label{eq:themetric2}
ds^{2} 
= 
f^{\, 2}(dt+\omega)^{2}
-f^{\, -1}d\hat{s}^{2}\, ,
\end{equation}
where the 1-form $\omega$ is obtained as
\begin{eqnarray}
\label{eq:omegahat2}
\omega 
& = & 
\omega_{5}(d\varphi+\chi) +\breve{\omega}\, ,
\\
& & \nonumber \\
\label{eq:omega52}
\omega_{5}
& = &
M+16\sqrt{2} H^{-2} C_{IJK} \Phi^{I} \Phi^{J} \Phi^{K}
+3\sqrt{2} H^{-1} L_I \Phi^{I} \, ,
\end{eqnarray}
being the inverse-hatted $\breve{\omega}$ the one in (\ref{eq:omegaequationB}), and $d\hat{s}^{2}$ is a four-dimensional Gibbons-Hawking metric \cite{Gibbons:1979zt,Gibbons:1987sp}

\begin{equation}
\label{eq:GHmetric2}
d\hat{s}^{2}
=
H^{-1} (d\varphi +\chi)^{2}
+H dx^{r}dx^{r}\, ,\,\,\,\, r=1,2,3\, .
\end{equation}

\item The physical vector fields and their field strengths are

\begin{equation}
\label{eq:vectorfields2}
\begin{array}{rcl}
A^{I} 
& = &
-\sqrt{3}h^{I} f (dt +\omega) +\hat{A}^{I}\, ,    
\\
& & \\
F^{I}
& = &
-\sqrt{3} \hat{\mathfrak{D}}[ h^{I} f (dt +\omega) ]  +\hat{F}^{I}\, ,
\end{array}
\end{equation}
where the auxiliary vectors $\hat{A}^{I}$ are four-dimensional gauge fields defined on the Gibbons-Hawking space as
\begin{equation}
\label{eq:instantondec2}
\begin{array}{rcl}
\hat{A}^{I}
& = &
-2\sqrt{6} \left[-H^{-1}\Phi^{I} (d\varphi+\chi)+\breve{A}^{I} \right]\, ,
\\
& & \\
\hat{F}^{I}
& = &
-2\sqrt{6}  
\left[-\breve{\mathfrak{D}} \left[ \Phi^{I} H^{-1} (d\varphi+\chi) \right]
+\star_{3} \breve{\mathfrak{D}} \Phi^{I} \right] \, ,
\end{array}
\end{equation}
By this construction, which is due to Kronheimer \cite{KronMT}, the field strength $\hat{F}^I$ is self-dual in the Gibbons-Hawking space, describing an instanton configuration intimately related to a lower dimensional static monopole.

Notice that $\hat{\mathfrak{D}}$  is the covariant derivative with associated connection $\hat{A}^{I}$ in the Gibbons-Hawking space, while $\breve{\mathfrak{D}}$  is the covariant derivative with associated connection $\breve{A}^{I}$ in $\mathbb{E}^3$.

\end{enumerate}
\end{enumerate}


\renewcommand{\leftmark}{\MakeUppercase{Bibliography}}
\phantomsection
\bibliographystyle{JHEP}
\bibliography{references}
\label{biblio}

\end{document}